\documentclass[[journal = nalefd, manuscript = article, layout = onecolumn]{achemso}
\usepackage{amsmath}
\usepackage{achemso}
\usepackage[T1]{fontenc}
\usepackage{xcolor}
\usepackage{ulem}
\def\_#1{\textsubscript{#1}}                                                                \def\^#1{\textsuperscript{#1}}

\author{Zachary J. Krebs} 
\altaffiliation{These authors contributed equally to this work.}
\author{Wyatt A. Behn} 
\altaffiliation{These authors contributed equally to this work.}
\author{Songci Li}
\author{Keenan J. Smith}
\affiliation{University of Wisconsin-Madison, Department of Physics, 1150 University Ave., Madison, Wisconsin 53706, USA}
\author{Kenji Watanabe}
\affiliation{
Research Center for Functional Materials,
National Institute for Materials Science, 1-1 Namiki, Tsukuba 305-0044, Japan}
\author{Takashi Taniguchi}
\affiliation{
International Center for Materials Nanoarchitectonics,
National Institute for Materials Science,  1-1 Namiki, Tsukuba 305-0044, Japan}
\author{Alex Levchenko}
\author{Victor W. Brar}
\affiliation{University of Wisconsin-Madison, Department of Physics, 1150 University Ave., Madison, Wisconsin 53706, USA}
\email{vbrar@wisc.edu}

\title{Imaging the breaking of electrostatic dams in graphene for ballistic and viscous fluids} 

\begin{document}

\begin{abstract}

The flow of charge carriers in materials can, under some circumstances, mimic the flow of viscous fluids.  In order to visualize the consequences of such effects, new methodologies must be developed that can probe the quasiparticle flow profile with nm-scale resolution as the geometric parameters of the system are continuously evolved.  In this work,  scanning tunneling potentiometry (STP) is used to image quasiparticle flow around engineered electrostatic barriers in graphene/hBN heterostructures.  Measurements are performed as \textit{electrostatic dams} --- defined by lateral pn-junction barriers --- are broken within the graphene sheet, and carriers move through conduction channels with physical widths that vary continuously from pinch-off to $\mu$m-scale.  Local, STP measurements of the electrochemical potential allow for direct characterization of the evolving flow profile, which we compare to finite-element simulations of a Stokesian fluid with varying parameters.  Our results reveal distinctly non-Ohmic flow profiles, with charge dipoles forming across barriers due to carrier scattering and accumulation on the \textit{upstream} side, and depletion \textit{downstream}.  Conductance measurements of individual channels, meanwhile, reveal that at low temperatures the quasiparticle flow is ballistic, but as the temperature is raised there is a Knudsen-to-Gurzhi regime crossover where the fluid becomes viscous and the channel conductance exceeds the ballistic limit set by Sharvin conductance. These results provide a clear illustration of how carrier flow in a Fermi fluid evolves as a function of carrier density, channel width, and temperature.  They also demonstrate how STP can be used to extract key parameters of quasiparticle transport, with a spatial resolution that exceeds that of other methods by orders of magnitude.

\end{abstract}

\section{Introduction}

The interactions between particles in a fluid play a critical role in determining the manner in which the fluid flows.  At low densities where particles can move ballistically, such as in gases, the conductance of a constriction is dependent on only the channel width and particle scattering from the walls, which leads to momentum loss.  At higher densities, particle-particle interactions - which preserve momentum - become more frequent and can lead to collective flows that enhance the conductivity through the constriction beyond the ballistic limit.\cite{knudsen1909law} This phenomenon, a behavior exhibited by viscous fluids with Laminar flow, was predicted by Gurzhi to also occur in electronic systems when the electron-electron (el-el) scattering length $l_{\text{ee}}$ becomes much shorter than momentum relaxing scattering lengths, $l_{\text{mr}}$.\cite{gurzhi1963minimum, gurzhi1968hydrodynamic} This behavior is observable in ultrapure material samples, and has been demonstrated through transport measurements in both PdCoO\_2\cite{moll2016evidence} and GaAs,\cite{de1995hydrodynamic} where other viscous, fluid-like behaviors have also been observed.\cite{braem2018scanning}
    
Recently, it has been predicted that such phenomena can also occur in graphene, where strong el-el interactions and low Umklapp scattering rates allow the quasiparticles to form viscous Fermi or Dirac fluids. \cite{guo2017higher, narozhny2017,lucas2018hydrodynamics, levitov2016electron, neto2009electronic,ho2018theoretical,principi2016bulk,torre2015nonlocal}   This fluid-like behavior could lead to several interesting phenomena to appear in graphene, including vortex formation, vortex shedding, and perhaps even electronic turbulence. \cite{mendoza2011preturbulent, levitov2016electron}  In order to observe these effects, a number of experimental methods have been implemented.  Transport measurements through a series of lithographed constrictions and strategically contacted samples have recently observed signatures of superballistic conductance as well as negative backflow - a possible indicator of vortex formation.\cite{berdyugin2019measuring,kumar2017superballistic,bandurin2018fluidity,bandurin2016negative}  Meanwhile, scanned single electron transistor and nitrogen vacancy measurements of etched, encapsulated graphene devices have imaged flow profiles with spatial resolutions as small as 50 nm, and observed signatures of Poiseuille flow, one potential sign of viscous flow behavior.\cite{sulpizio2019visualizing,ella2019simultaneous,jenkins2020imaging,ku2020imaging}  

In this work we use scanning tunneling potentiometry (STP) to image---with nm-scale spatial resolution---the electrochemical potential profile associated with quasiparticle flow in graphene around electrostatic barriers that are `drawn' using voltage pulses from the tip of the scanning tunneling microscope (STM).\cite{velasco_nanoscale_2016} This methodology allows for the creation of smooth barriers defined by in-plane p-n junctions which confine the particle flow without introducing diffusive scattering, or other momentum-relaxing processes which would occur in lithographed samples.\cite{kiselev2019boundary} Moreover, we are able to vary the width of the conduction channels from $\mu$m-scale to pinch-off (where the barriers form `electrostatic dams' that suppress flow). We also probe graphene/hBN samples that are non-encapsulated, which reduces charge screening, enhancing el-el interactions and allowing viscous flow behavior to be observable at shorter lengthscales.

Our results reveal how quasiparticle flow through constrictions changes as the carrier density, channel width, and temperature are varied.    We observe multiple signatures of non-Ohmic behavior, including a small drop in potential across the graphene, mean free paths that exceed 3 $\mu$m, and the formation of Landauer residual resistivity dipoles characterized by charge build-up and depletion on the upstream and downstream side of barriers, respectively.\cite{landauer1957spatial}  At 4.5 K, the flow behaves ballistically and we are able to observe ray-like streams of charge passing through the opened dam, with the channel conductance matching the Sharvin formalism.  Meanwhile, at 77 K, we observe a profile that more closely resembles viscous flow behavior, and we measure a channel conductance that is super-ballistic.  We find that our observations can be qualitatively-described by numerical simulations of a Stokesian fluid. We are able to estimate key parameters of the fluid, including $l_{ee}$, which we measure to be $\sim$100 nm at 77 K and a kinematic viscosity $\nu$ of $\approx 2.5 \times 10^3$ cm$^2$/s. These measurements represent a fundamentally new way of probing viscous electronic fluids that allows for controllable geometric effects on the flow to become observable.

\begin{figure}[t!]
    \centering
    \includegraphics[width
    =6.5in,keepaspectratio]{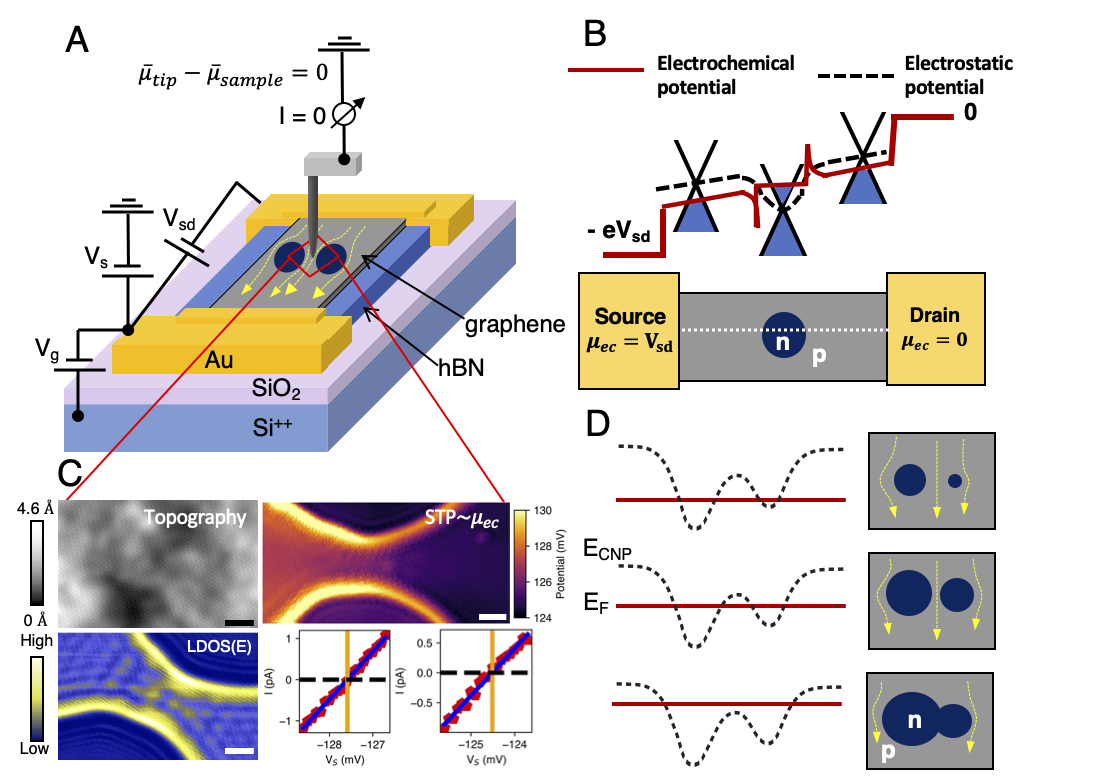}
    \caption{(A) Schematic of the STP experimental setup. $V_{sd}$ drives current in the sample while $V_s$ determines the difference between the sample and tip electrochemical potentials. The carrier density (and $E_F$) is globally modified through the use of an electrostatic gate electrode $V_g$. (B) A energy diagram depicting how the electrostatic potential and sample electrochemical potential vary along the flow direction across a potential well. (C) Simultaneously acquired topography [upper left] and spatial map of the electronic LDOS [lower left] ($V_s$ = -10 mV) of the electrostatic dam. All scale bars are 100 nm. Example STP map [upper right] ($V_g$ = -2 V) align with the current flow. Example tunneling I-V curves [lower right] acquired under transport. I = 0 (black dashed line) corresponds to $\mu_{ec} = -V_s$ (orange vertical line); by fitting many such curves (blue solid line) the sample electrochemical potential is mapped spatially. (D) Electrostatic dam shown for different gating conditions, from low to high electron doping (top to bottom). $E_{CNP}$ tracks the electrostatic potential in the graphene sheet (here shown moving across the channel).}
    \label{fig:overview}
\end{figure}

A schematic of our STP measurement geometry is shown in Fig. \ref{fig:overview}A.   In STP, a source-drain bias is used to drive current laterally through a thin sample, and the subsequent spatially varying electrochemical potential, $\mu_{ec}$, is measured locally using an STM tip.\cite{kirtley1988direct,briner1996local,chu1989scanning,muralt1987scanning,muralt1986scanning, druga2010versatile}  To measure $\mu_{ec}$, the feedback of the STM tip is turned off and the tip bias needed to zero the tunneling current is determined by performing a linear fit of the tip-sample I-V curve measured near the zero crossing (Fig. \ref{fig:overview}C).  This allows $\mu_{ec}$ to be measured with $\sim$ 10 $\mu$V potential resolution, and with the \AA-scale spatial resolution of standard STM. Fig. \ref{fig:overview}B illustrates how $\mu_{ec}$ varies across the sample under transport conditions.  When $V_{sd} = 0$, all local accumulations of charge which affect the chemical potential, $\mu$, are offset by changes in electrostatic potential $\phi$, such that $\mu_{ec}$ is constant across the surface. For $V_{sd} \neq 0$, meanwhile, $\mu_{ec}$ will change continuously across the sample, with a spatially varying slope that depends on the local conductance; meanwhile, any changes in local charge accumulation that are due to active carrier scattering or ballistic transport will affect $\mu_{ec}$ and be visible in STP measurement.\cite{bevan2014first,morr2017scanning} 

Previous STP measurements of graphene devices on SiC substrates have revealed sharp drops in potential associated with monolayer-bilayer boundaries, as well as sub-surface crystal steps.  In some cases, Landauer residual-resistivity dipoles could be observed near defect features, which could be used to model the electron-barrier scattering mechanisms.\cite{ji2012atomic,giannazzo2012electronic,wang2013local,willke2015spatial,clark2013spatially, sinterhauf2020substrate}  STP measurements have also been performed on graphene nanoribbons on SiC, where signatures of ballistic transport were observed.\cite{de2020non}  To the best of our knowledge, all previous STP measurements have been performed on SiC subtrates with large dielectric constants, which strongly screen el-el interactions.  Those measurements also utilized topographic features to act as scattering barriers, which are known to introduce artifacts in STP measurements due to tip convolution.\cite{pelz1990tip}

In this work, we probe ultraflat graphene/hBN samples with electrostatic barriers that are introduced by `drawing' them with the STM tip using a methodology developed by Velasco et al. \cite{velasco_nanoscale_2016, lee_imaging_2016, velasco2018visualization}. Each individual barrier was created by introducing sub-surface charges in the underlying hBN by applying a 1--2 minute, 5 V pulse with the STM tip which serves to ionize defects in the underlying hBN substrate.  Those defects create an electrostatic potential well in the plane of the graphene sheet that scatters incident holes and electrons. Within a suitable range of negative gate voltages (when the graphene is hole doped), a circular p-n (outside-inside) junction forms on the periphery of the potential well, which acts as a reflective boundary. By placing two of these p-n junctions in close proximity, we build a small channel that current can flow through when a source-drain bias is applied. Moreover, as shown in Fig. \ref{fig:overview}D, the width of this current-carrying channel can be tuned by using an electrostatic backgate to adjust the Fermi energy, $E_F$,  which alters the radius of the p-n junction barrier; the p-n junctions considered here decrease in radius with higher hole concentrations, which leads to an increased channel width.    

\begin{figure}[t!]
    \centering
    \includegraphics[width
    =6.3in,keepaspectratio]{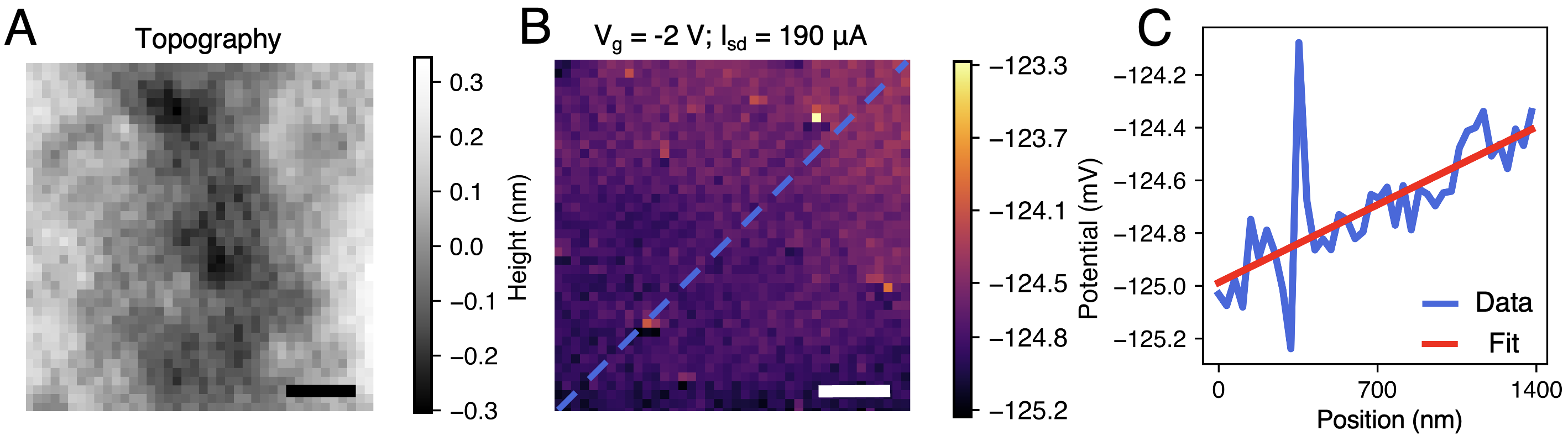}
    \caption{(A) Topographic STM image of a 1 x 1 $\mu m$ area of graphene on hBN. Scale bars are 200 nm. (B) Simultaneously acquired STP image from the same area obtained with $I_{\text{sd}}$ = 190 $\mu$A across a 30 $\mu m$ long sample that has an overall width of 15 $\mu m$. The periodic texture observed in both images is an aliasing effect created by the graphene/hBN Moire potential and the measurement grid. (C) Measured electrochemical potential along the flow direction and dashed line indicated in (B). The best fit line is shown in red (slope = 420 $\pm$ 10 uV/um).}
    \label{fig:slope}
\end{figure}

Prior to creating electrostatic barriers on our samples, we first obtain spatial maps of $\mu_{ec}$ when driving $I_{\text{sd}} = 190$ $\mu$A through the bare graphene/hBN sample, as shown in Fig. \ref{fig:slope}, along with a topographic image of the sample acquired simultaneously.  These data reveal a potential drop of 420 uV/$\mu m$ corresponding to a mean free path of $l_{\text{mr}} = 3$ $\mu$m at a low carrier density $n = -1.4 \times 10^{11}$ cm$^{-2}$, obtained via the Drude conductivity $\sigma = e^2 v_F l_{\text{mr}} D(E_F)/2$, where $v_F$ is the Fermi velocity and $D(E_F)$ the density of states at the Fermi level. \cite{sarma2011electronic} Some localized deviations in $\mu_{ec}$ are observable, which we attribute to charged defects buried in the hBN substrate.

\begin{figure}[t!]
    \centering
    \includegraphics[width
    =\textwidth,keepaspectratio]{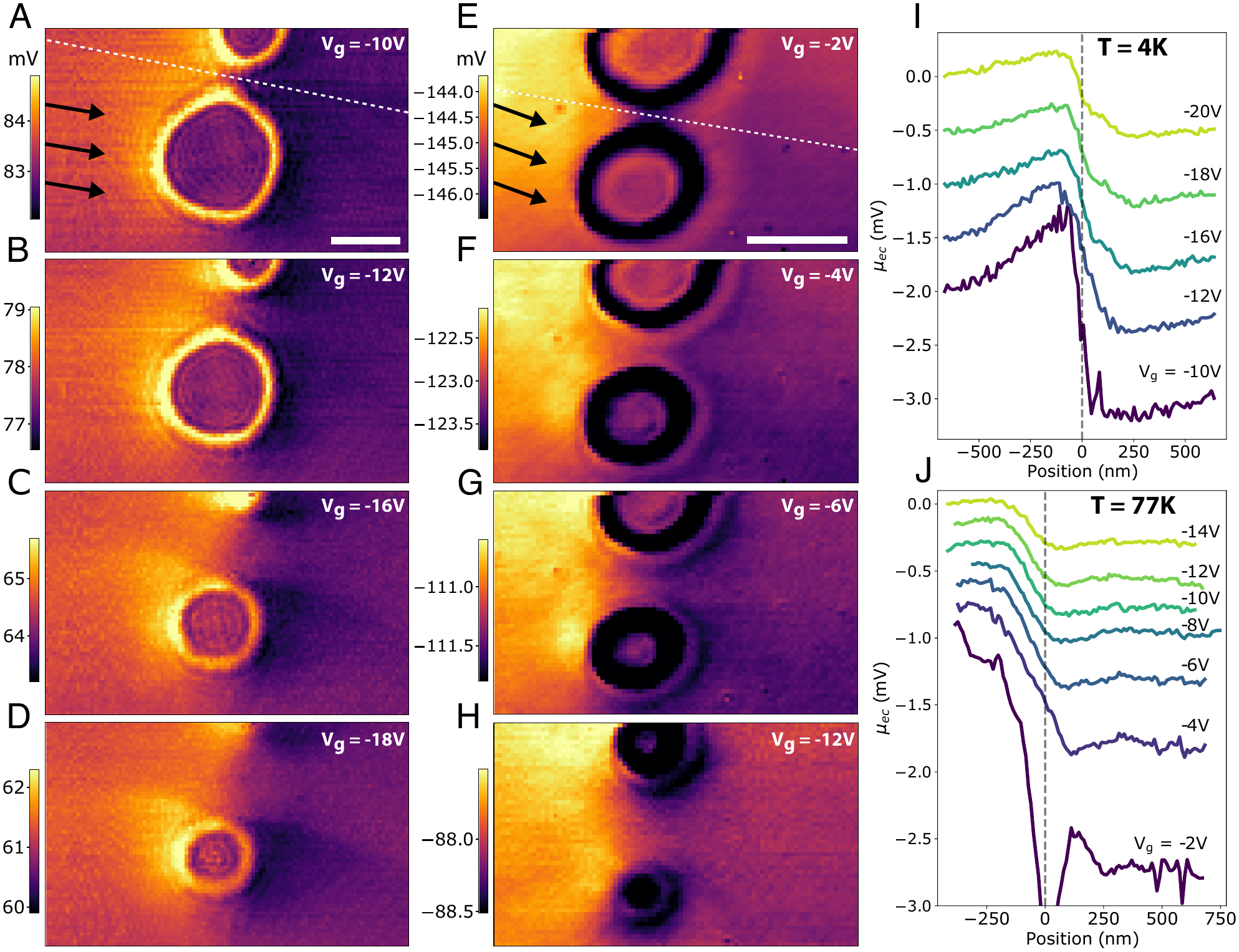}
    \caption{(A-D) STP maps of an electrostatic dam at T = 4.5 K, V$_{sd}=0.4$ V and four selected gate voltages: -10, -12, -16, and -18 V in order of increasing channel width. (E-H) STP maps of a new electrostatic dam at T = 77 K, V$_{sd}=-0.4$ V and four gate voltages: -2, -4, -6, and -12 V in order of increasing channel width. The scale bar is 250 nm. The black arrows represent the direction of current flow that is incident upon the barriers. (I) Line cuts through the STP maps at T = 4 K along the white dashed line in (A). (J) Line cuts through the STP maps at T = 77 K along the white dashed line in (E). Each curve is shifted by a constant offset for clarity. The dashed, vertical grey line marks the halfway point through the channels. }
    \label{fig:stp}
\end{figure}

STP images acquired after the formation of the potential wells at 4.5 K and 77 K are shown in Fig. \ref{fig:stp}, revealing a drastically altered electrochemical landscape. At both 4.5 K and 77 K, $\mu_{ec}$ is observed to increase (decrease) on the upstream (downstream) side of the potential wells, which creates in-plane dipoles across the wells. For the 4.5 K data, we associate these features with Landauer residual resistivity dipoles, which occur in ballistic (or near-ballistic) transport conditions when charge carriers scatter against localized potential barriers and accumulate (or are depleted) against the side of the barrier, which locally increases (decreases) the chemical potential.\cite{landauer1957spatial}  The in-plane dipole potentials decay approximately as $r^{-1}$, while the magnitude is determined by the current density.\cite{sorbello1981residual,sorbello1988residual}  We observe no significant changes in the cross-well dipole profile as the Fermi level of the device is changed by varying the electrostatic backgate voltage, $V_g$.  Within the quantum wells, standing waves associated with circular quasibound states that are excited by carriers are visible in the STP images due to their affect on the local charge density.\cite{bevan2014first}  The p-n barrier, meanwhile, can be observed as the bright (dark) ring in the 4.5 K (77 K) measurements.  The same ring feature has been determined in previous scanning tunneling spectroscopy and Kelvin probe force microscopy measurements to indicate the position of the classical turning point of the quasibound states, where there is an accumulation of quasiparticle density.\cite{gutierrez2018interaction,quezada2020comprehensive,velasco_nanoscale_2016, lee_imaging_2016,Behn_Krebs_2021}  It is not understood why the p-n barrier appears bright for measurements at 4.5 K, and dark at 77 K. This effect was observed over multiple, separate measurements, and we speculate that thermovoltages generated between the tip and the sample play an important role.\cite{sto1990thermopower,druga2010versatile,park2013atomic}   

In addition to the features described above, we also observe a drop in $\mu_{ec}$ along a transverse path through the channel between the wells, which represents the central focus of this work.  This change in $\mu_{ec}$ is associated with current that flows through the channel, the width of which can be tuned via electrostatic gating (as illustrated in Fig. 1D).  In Fig. \ref{fig:stp}, we show how the $\mu_{ec}$ landscape evolves as the channel is varied from as wide as 350 nm, to `pinch off' where it forms an `electrostatic dam', which blocks the incident current. These measurements are performed on separately prepared samples at $T = 4.5$ K (Fig. \ref{fig:stp}a-d) and at $T = 77$ K (Fig. \ref{fig:stp}e-h).  For 4.5 K measurements, ray-like `streams' of current are visible emerging from the downstream side of the channel, a property that is consistent with ballistic carriers passing through the gap, and locally increasing $\mu_{ec}$. \cite{morr2017scanning} Such qualitative `streams' are not as apparent for data obtained at 77 K.

\begin{figure}[t!]
    \centering
    \includegraphics[width
    =\textwidth,keepaspectratio]{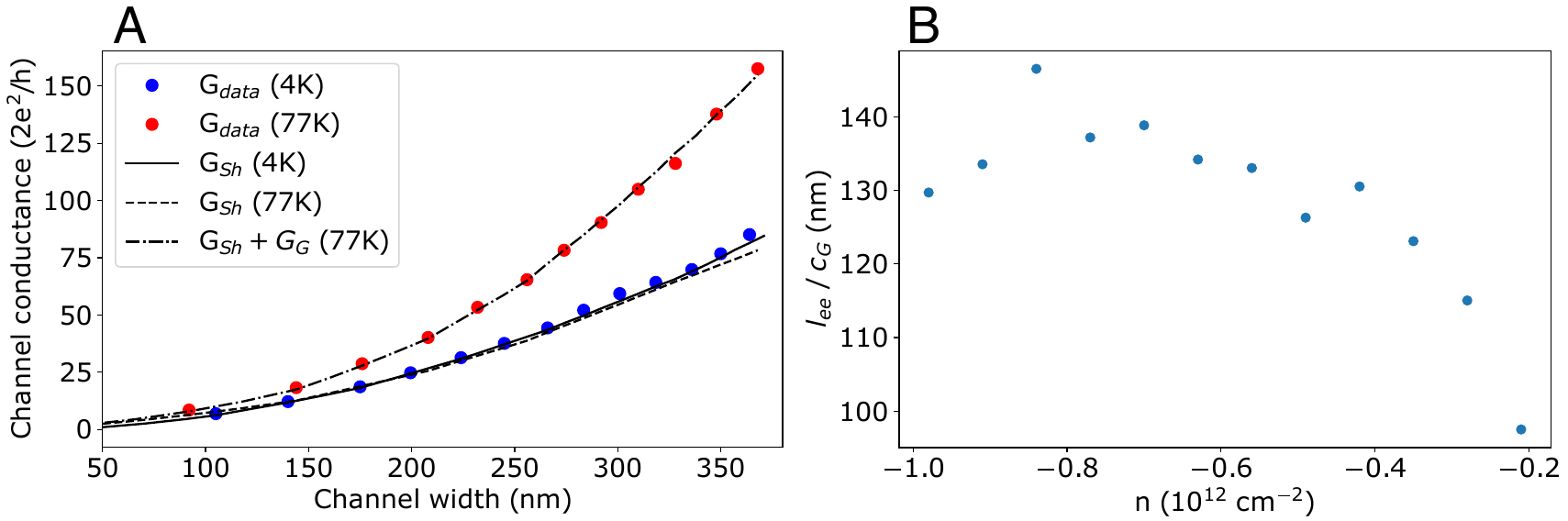}
    \caption{(A) Width-dependent channel conductance of electrostatically defined channels at 4.5 K and 77 K.  Solid (4.5 K) and dashed (77 K) lines indicate the theoretical ballistic conductance defined by the Sharvin formalism. (B) Carrier density-dependent electron-electron scattering lengths divided by the non-universal factor $C$, extracted from the superballistic conductance in (A).}
    \label{fig:conductances}
\end{figure}

In order to quantitatively characterize the carrier flow, we measure the width (and $V_g$)-dependent electrochemical drop through the channels (shown in Fig. \ref{fig:stp}i,j) to calculate the conductance of each channel, $G_{data} = I/\Delta \mu_{ec}$. The current $I$ flowing through each channel is first estimated using the rudimentary assumption that the ratio between the channel width and the width of the graphene flake (15 $\mu$m) is equal to the ratio of $I$ to the current passing through the whole flake, which is measured using a current meter.  Conductance values estimated in this way are compared against the values predicted by the Sharvin formula for ballistic transmission through a channel
\begin{equation}
    G_{\text{Sh}} = c_{\text{S}} G_{\text{Q}} \frac{\overline{E_F}}{\pi \hbar v_F} w
\end{equation}
where $G_{\text{Q}}=2e^2/\pi\hbar$, is the conductance quantum, $w$ is the width of the narrowest cross-section spanning the channel, $\overline{E_F}$ is the Fermi level (chemical potential) averaged over the the narrowest cross-section of the channel, and $c_{\text{S}}$ is a non-universal numerical factor specific to our electrostatic dams. Geometrical factors such as $c_{\text{S}}$ are known to effect current flows in both ballistic and viscous systems, and represent corrections of order 1 that are difficult to accurately calculate analytically. We determine $c_{\text{S}}$ by assuming that the transport at 4.5 K is ballistic, and therefore the conductivity depends linearly on $\overline{E_F}$ and $w$ for small channel widths as in Eq. (1). In repeated measurements at 4.5 K using a new electrostatic dam each time, we find excellent agreement between $G_{\text{Sh}}$ and $G_{data}$ across all channel widths when $c_{\text{S}} = 2.8$. The results for a selected measurement are shown in Fig. \ref{fig:conductances}. The fact that $c_S$ is greater than unity can be attributed to the details of the channel geometry and the finite extent of the electrostatic dams, as well as the nonzero transmissibility of the p-n junction barriers. We note that tip-induced doping effects and inhomogenous doping within the channel are unlikely to be significantly affecting these conductivity measurements, as $G_{data}$ taken with different tips and separately prepared potential barriers result in the similar values for both 4.5 K and 77 K. In contrast to measurements at 4.5 K, data taken at 77 K demonstrates a channel conductance that is \textit{larger} than the ballistic Sharvin model with $c_{\text{S}}=2.8$, and this deviation increases as the channel is widened. Using a larger number for $c_{\text{S}}$ in the 77 K theory does not produce good agreement across all channel widths. This suggests a viscous Gurzhi contribution to the channel conductance at elevated temperatures, $G_{\text{G}}$, which has a quadratic dependence on channel width $w$ and must be added to the Sharvin conductance $G_{\text{Sh}}$ to get the total conductance.  With this addition, we can estimate the electron-electron scattering length $l_{\text{ee}}$ that would lead to this enhanced conductance, we turn to \cite{guo2017higher, kumar2017superballistic}
\begin{equation}
    G = G_{\text{Sh}} + G_{\text{G}}, \quad G_{\text{G}} = c_{\text{G}} G_{\text{Q}} \frac{\overline{E_F}}{\pi \hbar v_F} \frac{w^2}{l_{\text{ee}}}
\end{equation}
from which we can write
\begin{equation}
    l_{\text{ee}} =w\, (c_{\text{G}}/ c_{\text{S}}) \left(\frac{G_{data}}{G_{\text{Sh}}} -1 \right)^{-1}
\end{equation}
where $c_{\text{G}}$ is an additional non-universal geometric numerical factor that is specific to viscous flow around our electrostatic dams. For example, $c_{\text{G}} = \pi^2/16$ for a perfect slit geometry \cite{guo2017higher}, however the value of $c_{\text{G}}$ also depends on the boundary conditions and differ for flows with no-slip and no-stress conditions.\cite{kiselev2019boundary,Pershoguba2020,Li2021}  Our resulting estimates of $l_{\text{ee}}$ are shown in Fig. \ref{fig:conductances}(b), giving values that are broadly consistent with previously published results.\cite{kumar2017superballistic} 

\begin{figure}[t!]
    \centering
    \includegraphics[width
    =6in,keepaspectratio]{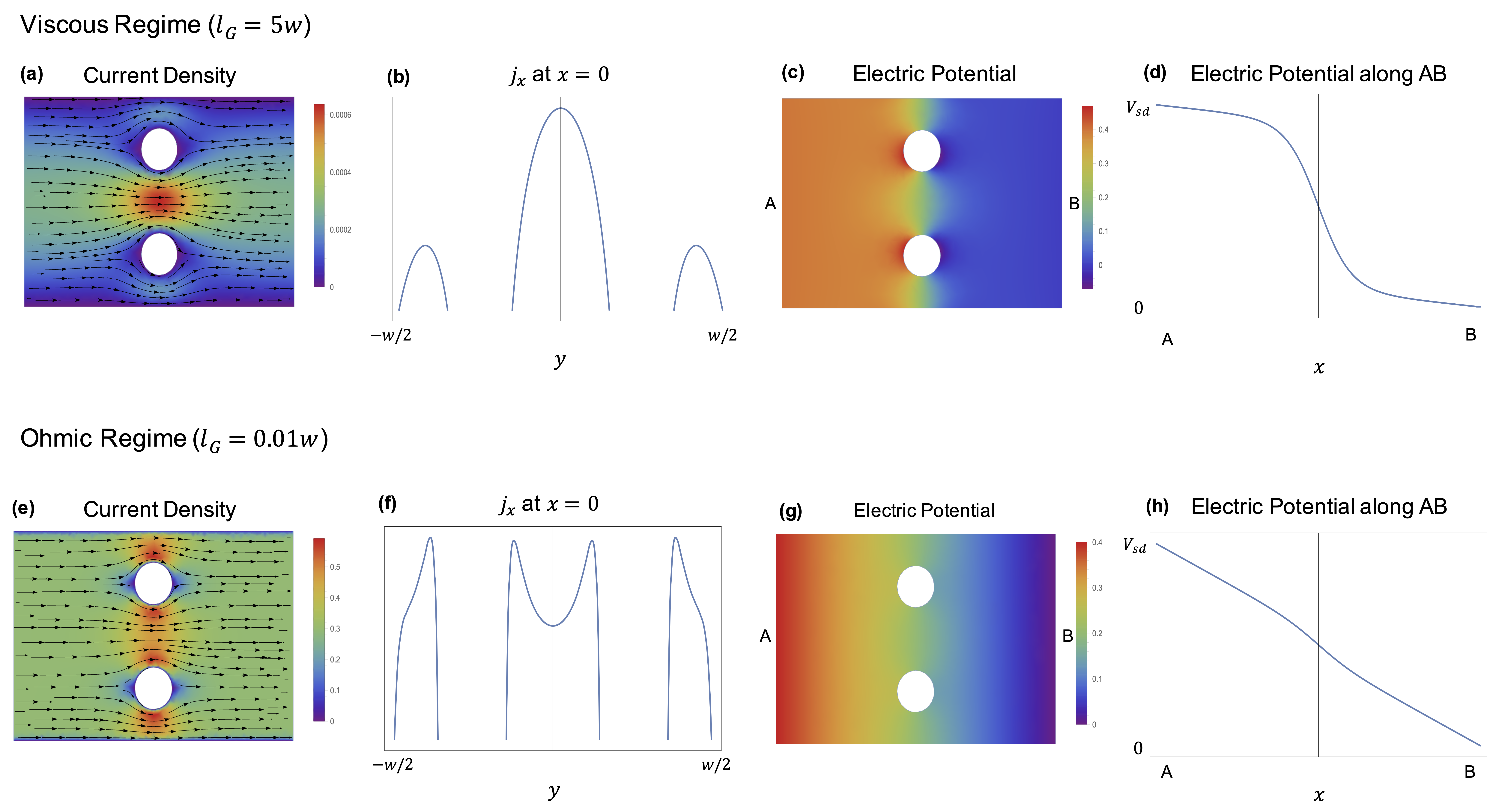}
    \caption{Finite element models of carrier flow in the Ohmic and viscous regimes. Numeric solutions of profiles of current density, electric potential for the hydrodynamic flow through two circular barriers in the viscous (upper panel) regime and Ohmic (lower regime) regime. $w$ is the width of the sample. (\textbf{a}) and (\textbf{e}): the arrow plots show the streamline of the current density and the color plots are the magnitude of the current density; (\textbf{b}) and (\textbf{f}): profiles of $j_x$ at $x=0$; (\textbf{c}) and (\textbf{g}): distribution of the electric potential; (\textbf{d}) and (\textbf{h}): line cuts of electric potential along $y=0$.}
    \label{fig:theory}
\end{figure}

These findings indicate that as the temperature of the graphene is increased from 4.5 K to 77 K, the el-el scattering length ($l_{ee}$) decreases until it is comparable to the width of the channel.  Under these conditions, the carrier flow transitions from a Knudsen to a Gurzhi regime, where it behaves as a viscous Fermi liquid, which exhibits a channel conductance that is greater than ballistic.   In order to better understand the potential profiles measured using STP and how they relate to viscous flow, we compare our measurements to the following theoretical model.  The motion of hydrodynamic electron flow under moderate external drive is can be described by the linear Navier-Stokes (NS) equation in the following form,
\begin{equation}
	\nu\nabla^2\mathbf{u}-\frac{\mathbf{u}}{\tau_{\text{mr}}}=\frac{e}{m}\nabla\phi, \label{eq:NS-1}
\end{equation}
where $\mathbf{u}$ is the macroscopic flow velocity, $\tau_{\text{mr}}$ is the 
momentum relaxation time and $\phi$ is the electric potential. It is evident that the first term on the left hand side of Eq.~\eqref{eq:NS-1} describes the viscous stress while the second term gives the Ohmic loss. In addition, the continuity equation for current conservation is written as
\begin{equation}
	\nabla \cdot \mathbf{j}=0. \label{eq:continuity}
\end{equation}
Considering that $\mathbf{j}=ne\mathbf{u}$ and assuming constant electron density $n$, the linear NS equation~\eqref{eq:NS-1} is recast in the form,
\begin{equation}
	l^2_{\text{G}}\nabla^2\mathbf{j}-\mathbf{j}=\sigma\nabla\phi, \label{eq:NS-2}
\end{equation}
where $\sigma$ is the Drude conductivity introduced earlier. The interplay of viscous and momentum relaxing terms introduces the natural length scale in the problem, namely the Gurzhi length, $l_{\text{G}}\equiv\sqrt{\nu\tau_{\text{mr}}}=\sqrt{l_{\text{ee}}l_{\text{mr}}}$. If $l_{\text{G}} \ll w$, where $w$ is the typical size of the system such as width of the channel, the viscous stress in Eq.~\eqref{eq:NS-2} can be neglected and one is in the Ohmic (diffusive) regime. In contrast, if $l_{\text{G}} \gg w$, the Ohmic dissipation in Eq.~\eqref{eq:NS-2} is small and one is in the viscous regime. 

In this framework, Eqs.~\eqref{eq:continuity} and~\eqref{eq:NS-2} lead to the profile of current density and electric potential. The analytical solutions are difficult to obtain for arbitrary geometries, nevertheless we provide numeric solutions for the hydrodynamic flow bypassing two circular barriers with finite element methods. The major results for the distribution of the current density and the electric potential are shown in Fig.~\eqref{fig:theory}. In the numeric simulation, we used no-slip boundary condition for the flow velocity $\mathbf{u}$. The flow is driven by the bias voltage $V_{\text{sd}}$, applied at the left side of the sample (the right side is grounded to zero). It is interesting to notice the dipole formation in the electric potential profile in the viscous regime, see Fig.~\eqref{fig:theory}(c). Such dipole formation or the increase of potential near the edges has been pointed out in Ref.~\cite{guo2017higher}, wherein the potential grows near the slits and diverge at the end points. This behavior can be attributed to the fact that the electric fields near the edges point against the current flow in order to push the electron liquid away from the boundary walls. It is also noteworthy that we have adopted hard wall potentials at the boundaries of the two circular barriers and neglected the nonlinear screening effects of the p-n interface. Accounting for these effects require solving self-consistent Poisson equation coupled with hydrodynamic flow equations~\eqref{eq:continuity} and~\eqref{eq:NS-2}, a simpler description would only involve a single circular p-n junction without considering the hydrodynamic flow. 
\begin{figure}[t!]
    \centering
    \includegraphics[width
    =\textwidth,keepaspectratio]{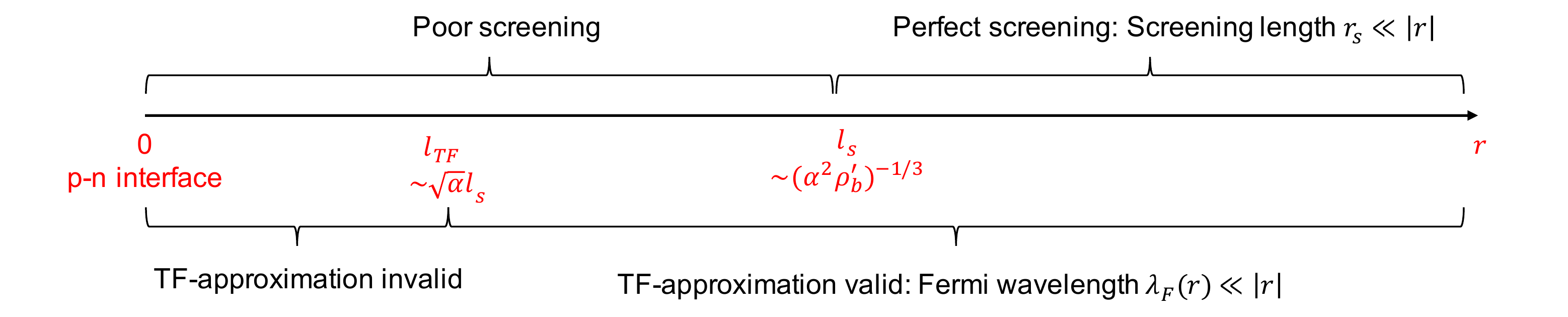}
    \caption{Regions of perfect screening and Thomas Fermi approximation. $\alpha=e^2/(\kappa\hbar v_F)$ is the interaction constant, $\kappa$ is the dielectric constant. $\rho'$ is the density gradient at the p-n interface.}
    \label{fig:screening}
\end{figure}

To assess the validity of electron flow modeling at constant density and nuances related to locally measured electrochemical potential, one can benefit from the analysis of nonlinear screening effects of a circular graphene pn-junctions that mimic our electrostatic barriers. The electric potential and charge density can be found self-consistently. We denote the background charge density created by external gates as $\rho_b(r)$, which can be approximated as $\rho_b(r)=\rho'_b\,r$ for the entire pn-junction width (the p-n interface is shifted to the origin $r=0$). The electric potential and the density can be solved from the following coupled equations,
\begin{subequations}
\begin{align}
    & \frac{\kappa}{e}V(r)=\int^\infty_0 \frac{4r'dr'}{r+r'}K\left(\frac{2\sqrt{rr'}}{r+r'}\right)\left[\rho_b(r')-\rho(r')\right], \label{eq:Poisson} \\
    & \mu[\rho(r)]-eV(r)=0,\quad \mu(\rho)=\sqrt{\pi}\hbar v_F\sqrt{|\rho|}, \label{eq:TF}
\end{align}
\end{subequations}
where $\kappa$ is the dielectric constant and $K(k)$ is the elliptic integral of the first kind with modulus $k$. In Eq.~\eqref{eq:TF}, we assumed the Thomas-Fermi (TF) approximation. It is then evident that within the TF approximation, the charge density satisfies the following equation:
\begin{equation} \label{eq:rho}
    \sqrt{\rho(r)}=\frac{\alpha}{\sqrt{\pi}} \int^\infty_0\frac{4r'dr'}{r+r'}K\left(\frac{2\sqrt{rr'}}{r+r'}\right)\left[\rho'_b r'-\rho(r')\right].
\end{equation}
The solution to Eq.~\eqref{eq:rho} gives the universal spatial profile of the charge density/potential, but it requires numerical evaluation. Nevertheless, one is still able to draw several qualitative conclusions from Eq.~\eqref{eq:rho}. Following similar analysis of a planar junction in Ref.~\cite{Fogler2008}, it can be shown that for small interaction constant $\alpha<1$, in the region $|r|\gg l_s\sim (\alpha^2 \rho'_b)^{-1/3}$, one has almost perfect screening, which can be described by TF approximation; in the region $l_{\text{TF}}\sim\sqrt{\alpha}l_s \ll |r|\ll l_s$, the screening effect is poor but the TF approximation still holds; in the immediate vicinity of p-n interface, $0<|r|<l_{\text{TF}}$, the TF approximation breaks down and one needs to compute the electron wavefunction in order to obtain the potential and the charge density. 
Given the material parameters, one can get the estimation $\alpha\approx 0.8$, $l_s\sim 116$nm, $l_{\text{TF}}\sim 103$nm (the junction width is $\sim 1\mu$m). The various domains and length scales are summarized in Fig.~\ref{fig:screening}. This means that almost everywhere except for a close proximity to the interface the macroscopic electron flow description gives reasonably accurate approximation.

In conclusion, we have shown that scanning tunneling potentiometry can be used to visualize hydrodynamic effects in graphene through direct imaging of the local electrochemical potential while a current is passed through the graphene sheet.  This methodology offers a superior spatial resolution to other scanned probe measurements, and allows for the creation and analysis of complex flow geometries defined by smooth barriers created by in-plane p-n junctions.  In this work, we show that STP can reveal super-ballistic conductance through narrow channels in graphene, as well as local dipoles that form in both ballistic and viscous regimes due to local carrier accumulation.  These results provide new insight into carrier transport in graphene, and provide a framework for analyzing more complex flow patterns that are engineered to exhibit exotic effects, such as, for example, non-reciprocal flow\cite{geurs2020rectification}, or to measure turbulence, which could occur at timescales and lengthscales that are inaccessible to other local probes.\cite{mendoza2011preturbulent}  \AA-scale images, meanwhile, could be used to visualize atomistic transport features that are predicted to occur along grain boundaries and near defects.\cite{bevan2014first}

\section{Acknowledgment}

Work by Z. J. K., S. L., K. J. S., and A. L. was supported by the U.S. Department of Energy (DOE), Office of Science, Basic Energy Sciences (BES) Program for Materials and Chemistry Research in Quantum Information Science under Award No. DE-SC0020313.  Work by W. A. B. and V. W. B. was supported by the Office of Naval Research under Award No. N00014-20-1-2356.  The authors gratefully acknowledge the use of facilities and instrumentation supported by NSF through the University of Wisconsin Materials Research Science and Engineering Center (No. DMR1720415). K.W. and T.T. acknowledge support from the Elemental Strategy Initiative conducted by the MEXT, Japan, Grant Number JPMXP0112101001,  JSPS KAKENHI Grant Number 19H05790 and JP20H00354.

\bibliography{bibliography}

\end{document}